\documentclass{article}

\usepackage{arxiv}

\usepackage{times}  
\usepackage{helvet} 
\usepackage{courier}  
\usepackage{graphicx} 
\usepackage{natbib}  
\usepackage{caption} 

\usepackage{amsmath,amssymb,amsfonts}
\usepackage{algorithmic}
\usepackage{textcomp}
\usepackage{xcolor}
\usepackage{comment}

\usepackage{multirow}
\usepackage{arydshln}
\usepackage{booktabs}
\usepackage{algorithm}
\usepackage{float}
\usepackage{url}

\title{Learning to Infer Unobserved Behaviors: Estimating User’s Preference for a Site over Other Sites}
\author{ 
	Atanu R Sinha$^*$ \\
	Adobe Research \\
	India \\
	\texttt{atr@adobe.com} \\
	\And
	Tanay Anand$^*$ \\
	Media and Data Science Research, Adobe \\
	India \\
	\texttt{tana@adobe.com} \\
    \And
	Paridhi Maheshwari \\
	Stanford University \\
	USA \\
	\texttt{1997.paridhi@gmail.com} \\
    \And 
	A V Lakshmy \\
	IIT Madras \\
	India \\
	\texttt{avlakshmy@gmail.com} \\
    \And
	Vishal Jain \\
	IIT Bombay \\
	India \\
	\texttt{vishalj409@gmail.com} \\
}

\begin{document}

\maketitle
%\title{To Whom to Recommend: Learning about Individual's Unobserved Behaviors}

\begin{abstract}
A site’s recommendation system relies on knowledge of its users’ preferences to offer relevant recommendations to them. These preferences are for attributes that comprise items and content shown on the site, and are estimated from the data of users’ interactions with the site. Another form of users’ preferences is material too, namely, users’ preferences for the site over other sites, since that shows users’ base level propensities to engage with the site. Estimating users’ preferences for the site, however, faces major obstacles because (a) the focal site usually has no data of its users’ interactions with other sites; these interactions are users’ unobserved behaviors for the focal site; and (b) the Machine Learning literature in recommendation does not offer a model of this situation. Even if (b) is resolved, the problem in (a) persists since without access to data of its users’ interactions with other sites, there is no ground truth for evaluation. Moreover, it is most useful when (c) users' preferences for the site can be estimated at the individual level, since the site can then personalize recommendations to individual users. We offer a method to estimate individual user’s preference for a focal site, under this premise. In particular, we compute the focal site's share of a user's online engagements without any data from other sites. We show an evaluation framework for the model using only the focal site's data, allowing the site to test the model. We rely upon a Hierarchical Bayes Method and perform estimation in two different ways - Markov Chain Monte Carlo and Stochastic Gradient with Langevin Dynamics. Our results find good support for the approach to computing personalized share of engagement and for its evaluation.

\end{abstract}



\def\thefootnote{*}\footnotetext{Authors contributed equally. The work was done while all authors were in Adobe.}

\section{Introduction}
%Recommendations of items and content to users are de rigueur in the online world. Machine Learning (ML) research in user modeling fuel recommendations and appear in topics such as search, recommendations, targeting, propensity scoring, among others~\cite{hidasi2015session,elkahky2015multi,zheng2016neural,covington2016deep,yoganarasimhan2020search,allenby1999dynamic,hiemstra2021advances, vardasbi2020inverse,karatzoglou2013learning,chen2016learning, lee2010improving,simester2020targeting}. The recommendation of items is predicated on the user, to whom to recommend? In selecting among users to recommend to, the ML user behavior models rely upon \textit{observed behavior} logs generated when an online \textit{focal} firm's (or, firm) customers or users (hereafter, users) interact with the firm's website and app. That is, 

%The learning of behaviors on an online \textit{focal} firm (or, firm) is predicated on 
%on a focal online firm (or, firm), leverage 
%\textit{observed behavior} data generated when its customers or users (hereafter, users) interact with the firm's website and app. 
%readily available for that firm. 
While users' observed behaviors inform the firm, it is also interested in the same users' \textit{unobserved behaviors}; that is, behaviors occurring on \textit{other} online firms.
%, but those data are unobservable to the focal firm. 
For example, the focal firm's elation at observing purchases on its site, is tempered by thinking that the same users are purchasing elsewhere as well. If the focal firm can infer the incidences of purchase of its users on other firms, which are unobservable by the focal firm, that allows the firm to learn the proportion of purchases users have with it versus with other firms. Accordingly, it can target users having lower proportions with offerings, and reward users with higher proportions. We offer a Hierarchical Bayes (HB) approach which learns to infer for the firm, each of its individual user's incidence with other firms, from the same data of observed behaviors the firm possesses.
%, without obtaining any data of unobserved behaviors. 

While the example above refers to purchases, the learning by the focal firm extends to incidences of other metrics such as visits, search, time spent, page views, dollar spent, etc. on other firms. Henceforth, \textit{engagement} refers to all such metrics, and forms two types. The incidences of \textit{observed engagement} are known to the firm from its observed behavior data. The incidences of \textit{unobserved engagement} (with other firms) are not known to the firm; which we want to learn. 
%, are to be learned from the same data. 
We express the two incidences into a single proportion, termed share of engagements, defined as the ratio of incidences with the firm versus incidences with other firms. Our learning approach infers share of engagements only with observed behavior data. Specifically and importantly, we learn \textit{model parameters for each individual user}, to infer \textit{Personalized Share of Engagements} for each user (hereafter, PSE). On a site, a user's PSE=0.23 for the engagement metric \textit{visits} means that the site receives 23\% of all visits the user makes to the site \textit{and} other sites. PSE is valuable to the firm to learn the degree of engagement each of its users has with the firm versus other firms, affording individualized targeting of offerings and messages.  
Arguably, proportions of engagement can be estimated by survey-sampling users from time to time, or, using one-off sample based panel study~\footnote{\url{https://www.numerator.com/infoscout-omnipanel}}. These approaches are obviously deficient due to the dependence on small samples, recall errors in surveys~\cite{couper2000web}, unavailability of these metrics for every time period, and thus do not form a reliable, consistent approach. Moreover, such aggregate level computations are not valuable for the focal firm to understand each of its users. 

The focal firm uses the log data of its own site or app. The logs of other firms are not shared with the focal firm for privacy, business intelligence and legal reasons. Laws such as General Data Protection Regulation in Europe and California Consumer Privacy Act also put additional protections against sharing~\cite{voigt2017eu,rothstein2019california}. 
This makes the research problem worthwhile since it calls for learning to infer unobserved engagements from observed engagements, a problem which has received less attention in ML data mining research. 

Our approach relies on a two part Hierarchical Bayes (HB) model. Part one posits a distribution of time between two successive engagements across both focal firm and other firms
%~\cite{morrison1981predicting}
, termed Inter Engagement Time (IET). The IET distribution yields epochs at which engagements occur on both the focal firm and the other firms' sites. We present two IET distributions, Erlang-2 and Exponential. %, commonly used in the literature~\cite{morrison1988generalizing,montgomery1970stochastic}.
In part two, we allocate engagements specifically to the focal firm, using a Markov model. Allocated engagements on the focal firm's site can be taken to the data of the focal firm, to estimate the model parameters. 
%A Hierarchical Bayes approach is used for estimation since individual level parameter estimates for PSE are sought. 
To show generality, we use two methods of estimation - Markov Chain Monte Carlo (MCMC) and Stochastic Gradient with Langevin Dynamics (SGLD). Without unobserved engagements' data and thus lack of ground truth, we introduce a general evaluation strategy for this type of problem. 
%The firm's data of engagements is randomly split into two portions - one portion is not exposed to the model, thus representing \textit{unobserved engagements}, and the other portion, representing \textit{observed engagements}, is used for estimating PSE. We now validate the learned PSEs from observed engagements, with the ground truth PSEs which come from comparing engagements in the observed portion with that of the unobserved portion. 

Our contributions are: 
\begin{itemize}
    \item Learning to infer unobserved behaviors from focal firm's own observed behavior data. 
    \item For each individual existing user measuring her personalized share of engagement with the focal firm versus that of other firms. 
    % \item Generalize from estimation of share of wallet on an offline one-off, credit card data~\cite{chen2012modeling} to estimation of PSE on typical online, clickstream data. 
    \item Introducing an evaluation strategy using only focal firm's own data. Evaluation within the firm's own data is necessary since other firms' data are not available to the firm. 
%    \item Provide a model-predictive performance comparison of two estimation methods, Markov Chain Monte Carlo (MCMC) and Stochastic Gradient with Langevin Dynamics (SGLD), in the context of a Hierarchical Bayes model.
    % \item Extend model specification requiring hand-curated additional data for covariates~\cite{chen2012modeling}, to data-driven covariates found in typical clickstream data.
\end{itemize}

\section{Relevant Literature}
%The two relevant lines of work are modeling IET and missing or incomplete information.

\subsection{Estimating Inter-Engagement Time}
Estimation of IET of users on a focal firm's site using logs is found for purchases in~\cite{guo2009multi}, while ~\cite{bhagat2018buy} models users' repeat purchase time intervals using statistical distributions. But, these papers restrict to a firm with observed data, without inferring unobserved behaviors. 
%Our estimate of inter-engagement time takes into account all sites while using data only on firm's site and using a modeling approach to account for unobserved behaviors' impact. Other work~\cite{allenby1999dynamic} models user inter-purchase times with the focal firm, using a Hierarchical Bayes model based on the generalized gamma distribution. 
Model for inter-purchase times of users on an online site to provide demand-aware recommendations is shown in~\cite{yi2017scalable}, and for timing for placement of privacy indicators on a site is found in~\cite{egelman2009timing}. Users’ visit frequency on different online sites is modeled using a negative binomial distribution~\cite{lee2001modeling}, while~\cite{fox2006hierarchical} uses a Tobit model. The proportional hazard model~\cite{seetharaman2003proportional} is also used to model inter-purchase time distributions. In addition, prediction of return time of users to a website is modeled using a proportional hazard model~\cite{kapoor2014hazard}, and using a semi-Markov model, which includes factors such as boredom~\cite{kapoor2015just}. A non-parametric neural network-based approach to model inter-arrival times is also proposed~\cite{chen2018multivariate}. Based on user panel data, they rely on observable users’ data on all sites, not only the focal site. Moreover, panel data use only a small sample, for a time period, and thus, cannot estimate PSE for every user of a firm, nor estimate for any time period. These works inform our choice of IET distributions, described later. 

\subsection{Modeling Missing or Incomplete Data}
The premise of no data of unobserved engagements is common,  distinguishing our work from substantial computer science literature on incomplete or missing data~\cite{sovilj2016extreme}, latent attributes~\cite{palla2012infinite} and others. These works do not consider our premise of unavailable data and do not learn about unobserved behaviors. One exception in a different literature is the estimation of share of wallet from offline, credit-card purchase data for a one-off dataset of single category using an HB approach~\cite{chen2012modeling}. HB models are constructed in a hierarchical manner and estimated with Bayesian methods. A good review is offered in~\cite{allenby2006hierarchical}. %Hierarchical modeling is used to address paucity of data and heterogeneity. 
Our use of HB overcomes the lack of adequate data at the individual user level. Bayesian estimation is often performed using Markov Chain Monte Carlo (MCMC) method~\cite{allenby2006hierarchical}, although stochastic gradient based methods for approximate Bayesian inference are making inroads to provide efficient computation~\cite{mandt2017stochastic}. Unlike MCMC, which uses full batch, these stochastic gradient based approaches use small-batch or mini-batch. 
%Several alternatives exist including  (SGLD), Stochastic Gradient Fisher Scoring, and others. 
Besides MCMC, we use Stochastic Gradient with Langevin Dynamics (SGLD)~\cite{welling2011bayesian}.

We follow~\cite{chen2012modeling} in the modeling approach. %and in the use of HB to overcome the obstacle of few observations for most users. 
%One of our two estimation methods, uses MCMC, in line with~\cite{chen2012modeling}. 
We depart significantly from~\cite{chen2012modeling}, by (i) using only log data of the firm, (ii) not using other datasets of externally obtained profiling and demographic information for features, nor any hand-curated data, (iii) introducing an evaluation strategy within firm's own data, (iv) using two distributions for IET, and (v) using a second estimation method, SGLD, which uses small batch training and overcomes MCMC's full batch requirement. The relevance of (ii) lies in the fact that many log data do not contain profiling and demographic information to preserve privacy. Without handcrafting, we show that log data is usable to estimate PSE, where engagement represents any online metric of relevance. We compare estimation results across MCMC and SGLD.

Although unrelated to our work, we note that unobserved behaviors are examined in the systems area~\cite{basile2019optimization,saives2015identification}. Modeling of users' behaviors is germane to data mining research in ML, spanning search, recommendations, targeting, etc.~\cite{hidasi2015session,elkahky2015multi,zheng2016neural,covington2016deep,hiemstra2021advances, vardasbi2020inverse,karatzoglou2013learning,chen2016learning, lee2010improving}. The goals and methods of these papers are very different from our paper. 

\section{Data}
The data comprise user level, behavior log for four months of an online merchant (or, focal firm) of electronics and entertainment products. Over the four months, different users visit the site and view several products categories. Some users have many visits, yet others have a handful of visits. Some users view a single product category, while others view many categories. Data of users having engagements (visits) on the site over four months are stitched by anonymized ID. The final input data have instances of engagement per user, for $1750$ users, across $4$ months ($121$ days). The descriptive statistics for the entire duration of data for each user are: Mean $62.5$, SD $31$ and Median $67$ days. Additionally, the user specific logs contain $3$ feature information for each user: loyalty status, offers received, and total number of purchases. 

Visits form the metric of engagement. Our assumption is that these users may also visit other sites who sell electronics and entertainment products. There are a wide range of other online firms that sell such products. We do not need to identify the set of other firms; all behaviors of the focal firm's users on those other firms' sites constitute unobserved engagements. Using data of only this single focal firm, we learn personalized share of visits of each of the focal firm's users. 
%That is, we compute, for each user of this focal firm, the proportion of visits that come to this focal firm. 
For this kind of problem, ground truth dataset is not available since data on unobserved engagements are not accessible to the focal firm. We overcome this obstacle by proposing a novel approach. We simulate ground truth within the data of the focal firm and show evaluation.

\section{Model}

We define \textit{all sites} as the set comprising the focal site and the other sites. %Intuition for the model goes as follows. 
%First, we postulate a probability distribution of engagements by each user $i$ across all sites.
%, following established model in the social science literature~\cite{morrison1988generalizing}. 
%A draw from this probability distribution yields number of engagements by $i$ across all sites. The question is: How to assign number of engagements to the focal site? We allocate, for each $i$, engagements to the site by a Markov model having two states - focal site and other sites. By combining the number of engagements across all sites with Markovian probability of engagement with the focal site, we are in a position to map the model allocated engagements on focal site to data of observed engagements on the focal site, for estimating model parameters. One further aspect is that although we seek parameter estimates for each $i$, engagement data on the focal site are inadequate for most users due to very few visits. Thus, to estimate parameters for each $i$, a Hierarchical Bayes approach is called for.
For ease of exposition, we provide a road map of the six steps involved in our modeling approach. (i) We postulate a distribution $F_i(.)$ of inter-engagement time (IET) in days, where IET is a random variable denoting number of days between successive engagements of $i$-th user across all sites. (ii) We allocate, for each $i$, engagements to the site by a Markov model having two states - focal site and other sites. Given $F_i(.)$, a Markov model computes the probability that an engagement by $i$ belongs to the focal site, yielding model based engagements for the focal site. (iii) We combine the number of engagements across all sites with Markovian probability of engagement with the focal site to obtain number of observed engagements on the focal site. That is, by combining IET for all sites with the Markov model, we obtain IET for the focal site. Now, parameter estimates are learned by mapping IET on focal site to data of observed engagements on the focal site. (iv) To derive individual-level estimates, parameters for each $i$ are modeled as functions of $i$'s features available on the focal site's data. A Hierarchical Bayes approach is used to cover for lack of adequate data for each user to estimate individual level parameters. For ecological reality of log data, we do not use features from any outside source as those are difficult to obtain and can not be stitched to individual users of the site. (v) Final outputs include IET across all sites, and PSE, for each $i$. For estimation, MCMC and SGLD are used. (vi) For validation, we introduce a simulated truth framework relying only on the site's actual data. 

\subsection{Inter-Engagement Time (IET)}
To model IET, we borrow from the established literature on %timing of purchases~\cite{morrison1981predicting} and that of 
arrival times for scheduling and queuing~\cite{li2007analysis,korenevskaya2019retrial}, and that of purchase timing~\cite{chen2012modeling}. We demonstrate our model using two alternative candidate probability distributions for IET. We define that the $i$-th user's IET follows Erlang distribution with shape parameter $s$ and scale $\beta_i$, given by:
\begin{equation}
    f_i(t;s,\beta_i)=\frac {\beta_i ^{s}t^{{s-1}}e^{{-\beta_i t}}}{(s-1)!}
\end{equation}
Later, we present evaluation in support of the distribution. The closed form solution of its mean is $s/\beta_i$. It also has a useful property that the sum of $k$ independent Erlang random variables with shape $s$ and the same scale is an Erlang random variable with shape $k*s$ and the same scale. We work with both the Erlang-2~\cite{chen2012modeling} and the Erlang-1 as the IET distributions. Erlang-1 is also known as the exponential distribution, which finds strong grounding as distribution of time between events~\cite{li2007analysis,korenevskaya2019retrial}. Notably, Erlang-2 and Erlang-1 are special cases of the Gamma distribution. The advantage of the Gamma distribution is it yields a family of distributions with various forms depending on the values of the shape and scale parameters. Thus, our choice of the two IET distributions to depict the approach come from a fairly general family of distributions. Validation experiments compare performance of these two distributions.   

\subsection{Markov Model}
On occasion $\tau$, a user engages either with the site or with other sites; i.e., a user can be in one of two states: [site, other sites]. On successive occasions, a user can move among these two states. On occasion $\tau$ she can belong to either state in [site, other sites] and on the next occasion she can move to either state in [site, other sites]. The transition among states follows a Markov model, mimicking a long tradition of its use to represent online interactions of users~\cite{gunduz2003web,kammenhuber2006web}. 
%Engagement at $\tau$ is influenced by whether engagement at $\tau-1$ occurred on focal site as opposed to other sites, in line with the use of Markov chain. 
Let, $\phi_{i}$ be the probability that $i$-th user who engages with the site in $\tau-1$ returns in $\tau$ to the site for engagement, and $\lambda_{i}$ be the probability that $i$-th user who engages with other sites in $\tau-1$ returns in $\tau$ to engage with the other sites. The resulting two-state Markov transition matrix is shown in Table~\ref{tab:trans_matrix}. We note that we do not impose any restriction on the magnitude of the probabilities of transition, but let the model estimate them from the data. 
The steady state probability of user $i$ engaging with the site gives $i$-th  user's PSE as:

\begin{gather}
    PSE_i \cdot \phi_i + (1 - PSE_i) \cdot \lambda_i = PSE_i; or,   
    PSE_i = \frac {\lambda_i}{1 + \lambda_i - \phi_i} 
\end{gather}

\begin{table*}[!h]
\centering
% \small
  \begin{tabular}{c|c|c|c}
    \toprule
    {} & {} & $\tau$ & $\tau$  \\  
    \midrule
    {} & {} & Focal site  & Other sites  \\ \hline 
    $\tau-1$ & Focal site  & $\phi_{i}$ & $1-\phi_{i}$ \\ \hline 
    $\tau-1$ & Other sites  & $\lambda_{i}$ & $1-\lambda_{i}$ \\
  \bottomrule
\end{tabular}

\caption{Markov Transition Probabilities for $i$-th user between two states - focal site and other sites}
\label{tab:trans_matrix}
\end{table*}

\iffalse
\begin{figure}[htbp]
\centerline{\includegraphics[width=0.35\columnwidth]{plots/markov_chain.png}}
\caption{Markov Transition Probabilities for $i$-th user between two states - focal site and other sites} 
\label{markovmodel}
\end{figure}
\fi 

\subsection{Combining Markov Model and IET}
To derive IET distribution of $i$-th user for the focal site, we combine $i$-th user's IET distribution $F_i(.)$ for all sites, with the Markov model which helps assign $i$-th user's visit to the focal site from her visits to all sites. If $k$ is the number of unobserved engagements between two observed engagements, the IET for the focal site is the sum of $k+1$ random variables drawn from $F_i(.)$, given by $f_i(t; 2(k+1) ,\beta_i)$.

Using the Markov model, we account for unobserved engagements by computing the probability of $k$ unobserved engagements between 2 observed ones as:
\begin{equation}
    Q_i(k, \phi_i, \lambda_i) = 
    \begin{cases}
      \hspace{1mm} \phi_i & \text{k = 0} \\
      \hspace{1mm} (1 - \phi_i)(1-\lambda_i)^{k - 1}\lambda_i & \text{k} > 0 \\
    \end{cases}
\end{equation}

The distribution $g_{1i}(.)$ of $i$-th user's IET on focal site is obtained by summing over all $k$ (a large number for estimation):
\begin{equation}
    g_{1i}(t;\beta_i, \phi_i, \lambda_i) = \sum_{k=0}^{\infty} f_i(t; 2(k+1), \beta_i) \cdot Q_i(k, \phi_i, \lambda_i)
\end{equation}
Using the expectation of Erlang distribution with shape $s$ and scale $\beta_i$, it can be shown that the expected value of $g_{1i}(.)$, i.e, the expected value between observed engagements, is given by
\begin{equation}
    \frac{s}{\beta_i} \cdot \frac{1 + \lambda_i - \phi_i}{\lambda_i} = \frac{s}{\beta_i} \cdot \frac{1}{PSE_i}
\end{equation}

Parameters of the distribution $g_{1i}(.)$ are now estimable from data as engagements on the focal site are observed. Next, estimation of ($\beta_i$,\hspace{1mm}$\phi_i$,\hspace{1mm}$\lambda_i$), for each $i$ is described.

\subsection{Hierarchical Bayes Approach}
To estimate the above parameters directly, for each $i$, the constraint is that the number of data points per user (observed engagements) is not large enough, for most users. A Hierarchical Bayes approach overcomes the constraint. The hierarchy comes through by setting the prior distribution of ($\beta_i$,\hspace{1mm}$\phi_i$,\hspace{1mm}$\lambda_i$) to depend upon other parameters, with their own prior distribution. As shown below, the individual level parameters ($\beta_i$,\hspace{1mm}$\phi_i$,\hspace{1mm}$\lambda_i$), are expressed as functions of other parameters ($\boldsymbol{\eta}$, $\boldsymbol{\gamma}$, $\boldsymbol{\delta}$), common across individuals, that can be estimated using features, on which data are available for each $i$. First, we reparameterize ($\beta_i$,\hspace{1mm}$\phi_i$,\hspace{1mm}$\lambda_i$) to impose desirable properties: $\beta_i > 0$, and 0 $<$ {$\phi_i$, $\lambda_i$} $<$ 1.
\begin{equation}
    \mathbf{\beta_i = \exp(\theta_{\beta_i}) \hspace{4mm}
    \phi_i = \frac {\exp(\theta_{\phi_i})}{1 + \exp(\theta_{\phi_i})} \hspace{4mm}
    \lambda_i = \frac {\exp(\theta_{\lambda_i})}{1 + \exp(\theta_{\lambda_i})}}
\end{equation}

Then ($\theta_{\beta_i}$,\hspace{1mm}$\theta_{\phi_i}$,\hspace{1mm}$\theta_{\lambda_i}$) are specified as functions of features from log data. Let, $\mathbf{X_{\beta_i}}$, $\mathbf{X_{\phi_i}}$ and $\mathbf{X_{\lambda_i}}$, denote three features: offers, loyalty and total number of purchases made on focal site. 
%These are features available from clickstream dataset itself, and the features are not informed by any other dataset. 
In a difference from~\cite{chen2012modeling}, which make use of other supplementary, hand-curated information obtained from outside resources, we do not. To make the model widely applicable for typical log data we refrain from using supplemental information. Linear regression model for each parameter is specified as:
% \begin{align} \label{linear}
%     \mathbf{\left(\begin{array}{c}{\theta_{\beta_i}} \\ {\theta_{\phi_i}} \\ {\theta_{\lambda_i}}\end{array}\right)
%     & = \left(\begin{array}{c}{\mathbf{X_{\beta_i}^{T}} \boldsymbol{\eta}} \\[0.1cm] {\mathbf{X_{\phi_i}^{T}} \boldsymbol{\gamma}} \\[0.1cm] {\mathbf{X_{\lambda_i}^{T}} \boldsymbol{\delta}}\end{array}\right)
%     + \left(\begin{array}{c}{\varepsilon_{\beta_i}} \\[0.1cm] {\varepsilon_{\phi_i}} \\[0.1cm] {\varepsilon_{\lambda_i}}\end{array}\right)} \\[0.25cm]
%     \mathbf{\Theta_{i} & = \mathbf{A_{i}B}+\boldsymbol{\epsilon_i}},  
%     \text{where} \mathbf{\epsilon_{i}} \sim \mathbf{Normal(0, \Omega)} \nonumber
% \end{align}

\begin{align} \label{linear}
    \left(\begin{array}{c}
        \theta_{\beta_i} \\ 
        \theta_{\phi_i} \\ 
        \theta_{\lambda_i}
    \end{array}\right)
    &= \left(\begin{array}{c}
        \mathbf{X_{\beta_i}^{T}} \boldsymbol{\eta} \\[0.1cm] 
        \mathbf{X_{\phi_i}^{T}} \boldsymbol{\gamma} \\[0.1cm] 
        \mathbf{X_{\lambda_i}^{T}} \boldsymbol{\delta}
    \end{array}\right)
    + \left(\begin{array}{c}
        \varepsilon_{\beta_i} \\[0.1cm] 
        \varepsilon_{\phi_i} \\[0.1cm] 
        \varepsilon_{\lambda_i}
    \end{array}\right) \\[0.25cm]
    \mathbf{\Theta_{i}} &= \mathbf{A_{i}B} + \boldsymbol{\epsilon_i}, \nonumber \\
    & \text{where } \boldsymbol{\epsilon_{i}} \sim \mathbf{Normal(0, \Omega)} \nonumber
\end{align}
with $\mathbf{A_i}$ as block diagonal matrix, blocks refer to ($\mathbf{X_{\beta_i}}$, $\mathbf{X_{\phi_i}}$ and $\mathbf{X_{\lambda_i}}$), $\mathbf{B}$ is parameter column vector ($\boldsymbol{\eta}$, $\boldsymbol{\gamma}$, $\boldsymbol{\delta}$) and $\boldsymbol{\epsilon_i}$ follows a multivariate normal distribution. 

To recognize that PSE is heterogeneous across individuals, we draw $PSE_i$ from normal distribution ${g_{2i}(PSE_i|\mathbf{\Theta_{i}})}$, such that, 
\begin{gather}
    \frac{PSE_{i}}{1-PSE_{i}} \sim \mathbf{Normal (\mu, \sigma^{2})}, 
    \text{where} \quad \mu = \frac{PSE_{agg}}{1-PSE_{agg}} 
\end{gather}
$PSE_{agg}$ may be obtainable from available market reports by the likes of Nielsen, ComScore, and Infoscout~\cite{omnipanel}. Such report can be available from a market research firm as a one time study and is not available perennially. The overall likelihood function for observed engagements is thus expressed as,
\begin{equation} \label{likelihood1}
    \prod_{i=1}^{n} \mathcal{L}_i \hspace{1mm} \text{where} \hspace{1mm} \mathcal{L}_i = { {\left( \prod_{j=1}^{m_i} g_{1i}\left(t_{i j} | \mathbf{\Theta_{i}}\right) \right)} \hspace{1mm} {g_{2i}(PSE_i|\mathbf{\Theta_{i}})} }\\
\end{equation}
where $n$ is number of users, and for each $i$, $\mathcal{L}_i$ is the likelihood, $t_{ij}$ is the $j$-th IET for observed engagements, and $m_i$ is number of IETs for observed engagements. 

A major question arises when $PSE_{agg}$ is not obtainable from market reports, or, it may be error prone. This situation is worthy of study to make the case that the focal site avoid the use of any data of other sites. Thus, in a concurrent examination, we ignore that $PSE_i$ follows a normal distribution ${g_{2i}(PSE_i|\mathbf{\Theta_{i}})}$. The likelihood function reduces to: 
\begin{equation} \label{likelihood2}
    \prod_{i=1}^{n} \mathcal{L}_i \quad \text{where} \quad \mathcal{L}_i = { \prod_{j=1}^{m_i} g_{1i}\left(t_{i j} | \mathbf{\Theta_{i}}\right) }
\end{equation}
In experiments we compare whether and how the use of aggregate market report based $PSE_{agg}$ and the use of consequent distribution ${g_{2i}(PSE_i|\mathbf{\Theta_{i}})}$ impact performance results.

% We present two methods of estimation, MCMC which relies on full batch training, and SGLD which uses a small batch to train. Later, we offer a head to head comparison in predictive performance of these two methods in estimating IET and PSE. 
% Details of both estimation methods and pseudo codes are available in the Appendix. 

\begin{table*}[!h]
\centering
\begin{tabular}{ ccc cc c cc c cc c cc }
    \toprule
    \multirow{2}{*}{\textbf{g2}} & \multirow{2}{*}{$n$} & \multicolumn{2}{c}{\textbf{MCMC Erlang-2}} && \multicolumn{2}{c}{\textbf{MCMC Erlang-1}} && \multicolumn{2}{c}{\textbf{SGLD Erlang-2}} && \multicolumn{2}{c}{\textbf{SGLD Erlang-1}} \\
    \cmidrule{3-4}\cmidrule{6-7}\cmidrule{9-10}\cmidrule{12-13}
    & & \textbf{RMSE} & \textbf{sMAPE} && \textbf{RMSE} & \textbf{sMAPE} && \textbf{RMSE} & \textbf{sMAPE} && \textbf{RMSE} & \textbf{sMAPE}\\
    \midrule
    Y & 1750 & 3.39 & 7.30\% && 1.92 & 3.77\% && 5.79 & 18.93\% && 7.50 & 19.59\% \\
    N & 1750 & 3.40 & 7.32\% && 3.81 & 12.47\% && 5.60 & 17.47\% && 6.01 & 17.76\% \\
    \bottomrule
\end{tabular}
\caption{Experiment 1 - Interim Evaluation of IET for focal site. RMSE and sMAPE values for g2=Y are comparable to that of g2=N. MCMC yields better model performance than SGLD, across both Erlang-2 and Erlang-1, as well as, for g2=Y and g2=N.}
\label{originalResults}
\end{table*}

\section{Estimation Algorithms}
We present two methods of estimation, MCMC which relies on full batch training, and SGLD which uses a small batch to train. Later, we offer a head to head comparison in performance of these two methods in estimating IET and PSE.

\subsection{Markov Chain Monte Carlo}
For the MCMC method, the Metropolis Hastings algorithm along with Gibbs Sampling is used. Normal distributions are used as priors due to its self-conjugate property. In each iteration, draws are generated from conditional posterior distributions, first for $\mathbf{\Theta_i}$, and then, conditional on $\mathbf{\Theta_i}$, for ($\boldsymbol{\eta}$, $\boldsymbol{\gamma}$, $\boldsymbol{\delta}$) and $\boldsymbol{\Omega}$. In each iteration, a pass is made over all $n$ observations.

We start by randomly sampling $\mathbf{\Theta_i}$ from its conditional posterior distribution, which uses equation \ref{likelihood1} and is shown below.
\begin{gather} \label{f}
    \mathbf{f\left(\Theta_{i}\right)} \propto  \mathbf{|\boldsymbol{\Omega}|^{-\frac{1}{2}} \exp {\left[ -\frac{1}{2} \left(\Theta_{i}-\overline{\Theta}_{i}\right)^{T} \Omega^{-1}\left(\Theta_{i}-\overline{\Theta}_{i}\right)\right]}} \mathcal{L}_i
\end{gather}
The generated sample for $\mathbf{\Theta_i}$ is updated as 
\begin{align*}
    \mathbf{\Theta_{i}}^{(l)} & = \mathbf{\Theta_{i}}^{(l-1)}+\Delta, 
    \text{where   } \Delta  \sim \mathbf{Normal}(0, scale)    
\end{align*}
here, $scale$ is set as a hyperparameter. The $l$-th updated value is accepted if a random number between $\mathbf{0}$ and $\mathbf{1}$ is less than the quantity, 
\begin{equation} \label{acceptance condition}
    \min \left[ \hspace{1mm} \frac{\mathbf{f}\left(\mathbf{\Theta_{i}}^{(l)}\right)}{\mathbf{f}\left(\mathbf{\Theta_{i}}^{(l-1)}\right)} \hspace{1mm},\hspace{1mm}1\hspace{1mm} \right]
\end{equation}
else, we reject the update and retain the $(l-1)$-th value.

Conditional on updated values of $\mathbf{\Theta_{i}}$, new values of ($\boldsymbol{\eta}$, $\boldsymbol{\gamma}$, $\boldsymbol{\delta}$) and $\boldsymbol{\Omega}$ are generated using Gibbs sampling. The prior distribution for ($\boldsymbol{\eta}$, $\boldsymbol{\gamma}$, $\boldsymbol{\delta}$) is a multivariate normal given by,
\begin{align*}
    \mathbf{P(\eta, \gamma, \delta) \sim Normal(0, 100I)}
\end{align*}
where $\mathbf{I}$ is the identity matrix. The posterior distribution for ($\boldsymbol{\eta}$, $\boldsymbol{\gamma}$, $\boldsymbol{\delta}$) is conditional on updated values of $\mathbf{\Theta_{i}}$ and the current values of $\boldsymbol{\Omega}$
and is sampled from a multivariate normal which can be obtained by rearranging the terms as follows,
\begin{gather}
    \mathbf{B = A_{i}^{+} \, (\Theta_{i} - \epsilon_{i})},  
    \text{where  } \mathbf{A_{i}^{+} = (A_i^{T} A_{i}^{})^{-1} A_i^{T}} \nonumber
\end{gather}
here $\mathbf{A_{i}^{+}}$ denotes the pseudo inverse of $\mathbf{A_{i}}$. While the prior distribution of $\boldsymbol{\Omega}$ is an Inverse Wishart distribution, \begin{align*}
    \mathbf{P(\Omega) \sim \mathcal{W}^{-1}(\mathbf{\Psi}, \nu)}
\end{align*}
and the posterior distribution of $\boldsymbol{\Omega}$, conditional on updated values of $\mathbf{\Theta_{i}}$ and ($\boldsymbol{\eta}$, $\boldsymbol{\gamma}$, $\boldsymbol{\delta}$), with $\mathbf{X_{\epsilon} = \Theta_{i}-\overline{\Theta}_{i}}$, is given by,
\begin{align*}
    \mathbf{P(\Omega\hspace{1mm}|\hspace{1mm}\Theta, \boldsymbol{\eta}, \boldsymbol{\gamma}, \boldsymbol{\delta)}} \sim \mathcal{W}^{-1}(\mathbf{X_{\epsilon}X_{\epsilon}^T + \Psi}, n + \nu)
\end{align*}
The MCMC is run for 30,000 iterations and the first 10,000 iterations are used as burn in. The results presented come from the last 20,000 iterations. We also tried runs of 50,000 iterations and using their last 20,000 iterations, but did not see significant change in convergence or results.

\begin{algorithm}
\caption{MCMC}
\begin{algorithmic}[1]
	\FOR {$iteration=1,2,\ldots,N$}
		\FOR {$user=1,2,\ldots,n$}
			\STATE $\mathbf{\Theta_{i}}^{(new)}$ $\leftarrow$ $\mathbf{\Theta_{i}}$ + $\Delta$ , where $\Delta \sim \mathbf{Normal}(0, scale)$
			\STATE $\mathbf{\Theta_{i}}$ $\leftarrow$ $\mathbf{\Theta_{i}}^{(new)}$ if acceptance condition \eqref{acceptance condition} is true
			\STATE $\mathbf{B}$ $\leftarrow$ $\mathbf{A_{i}^{+} \, (\Theta_{i} - \epsilon_{i})}$
		\ENDFOR
	\ENDFOR
	\STATE $(\beta_{i}, \phi_{i}, \lambda_{i})$ $\leftarrow$ reparameterize($\mathbf{\Theta_{i}}$)
\end{algorithmic} 
\end{algorithm}

\subsection{Stochastic Gradient Langevin Dynamics}
The second method used for parameter estimation is the Stochastic Gradient with Langevin Dynamics algorithm~\cite{welling2011bayesian}. Unlike in MCMC, we do not have an accept reject condition in SGLD. We do not sample ($\beta_i$,\hspace{1mm}$\phi_i$,\hspace{1mm}$\lambda_i$) directly, we instead compute them by plugging in co-variate values and estimated ($\boldsymbol{\eta}$, $\boldsymbol{\gamma}$, $\boldsymbol{\delta}$) in the linear equation~\ref{linear}. For estimation, we use the standard update step~\cite{welling2011bayesian} given by, 
\begin{align*}
    %\hspace{5mm} 
    \boldsymbol{\zeta_{l+1}} = \boldsymbol{\zeta_{l}} + \boldsymbol{\rho_{l}} + 
    %\hspace{10mm} 
    \frac{\tau}{2}\left\{\nabla_{\boldsymbol{\zeta_{l}}} \log p\left(\boldsymbol{\zeta_{l}}\right) + \frac{n}{n'} \sum \nabla_{\boldsymbol{\zeta_{l}}} \log f\left(\Theta_{i} \mid \boldsymbol{\zeta_{l}}\right)\right\} \\
    \text{where  } \boldsymbol{\zeta} = (\boldsymbol{\eta}, \boldsymbol{\gamma}, \boldsymbol{\delta}) 
    \text{and  } \mathbf{\boldsymbol{\rho_{l}} \sim Normal(0, \varepsilon_{l}\times\Omega)} \hspace{20mm}
\end{align*}
where summation is computed over all samples in a batch, $n'$ is the small batch size and $n$ is the total number of users. We have used a constant stepsize $\tau$ across iterations for the estimation.

The individual likelihood $f(\Theta_{i} \mid \boldsymbol{\zeta_{l}})$ is computed using equations~\ref{likelihood1} and ~\ref{likelihood2}. The prior distribution for $\boldsymbol{\zeta_{l}}$, that is $\mathbf{P(\eta, \gamma, \delta)}$, is a multivariate normal given by,
\begin{align*}
    \mathbf{P(\eta, \gamma, \delta) \sim Normal(0, 100I)}
\end{align*}

While $\boldsymbol{\Omega}$ is sampled from an Inverse Wishart distribution as,
\begin{align*}
    \mathbf{P(\Omega) \sim \mathcal{W}^{-1}(\mathbf{\Psi}, \nu)}
\end{align*}

We use a batch size of 200 and run for 30,000 iterations and the first 20,000 iterations are used as burn in. Iterations more than 30,000 give similar results.

\begin{algorithm}
\caption{SGLD}
\begin{algorithmic}[1]
	\FOR {$iteration=1,2,\ldots$}
		\FOR {$actor=a,a+1,\ldots,a+batchsize$}
			\STATE gradient-individual $\leftarrow$ Compute gradient of individual likelihood w.r.t $(\eta, \gamma, \delta)$
			\STATE gradient-mini-batch += gradient-individual
		\ENDFOR
		\STATE log-gradient $\leftarrow$ Compute gradient of prior of $(\eta, \gamma, \delta)$ w.r.t $(\eta, \gamma, \delta)$ 
		\STATE update-value $\leftarrow$ Compute update value using log-gradient and gradient-mini-batch using equation
		\STATE $(\eta_{t+1}, \gamma_{t+1}, \delta_{t+1})$ $\leftarrow$ $(\eta_{t}, \gamma_{t}, \delta_{t})$ + update-value
	\ENDFOR
\end{algorithmic} 
\end{algorithm}

\section{Experiments and Results}
Our dataset shows the metric, visits, to the focal site. In this implementation, thus, we estimate PSE for each user's visits, as personalized share of visits. The proposed PSE can apply to any metric, subject to availability of data for that metric for the site. 
%We curate the dataset using all visits for all categories made on the site. %While it is useful to estimate the model for individual categories, we refrain from doing so because of paucity of data for a single product category. 
Besides PSE, the model yields IET for each user. We devise $3$ different experiments for evaluation - one interim evaluation, and two validation experiments based on simulated data constructed from the site's own data. 

In each experiment, we show results using:
(a) two estimation algorithms - MCMC and SGLD; 
(b) two IET distributions - Erlang-2 and Erlang-1;
(c) two objective functions shown in equations \ref{likelihood1} and \ref{likelihood2}.
As a recall, likelihood in equation \ref{likelihood2}, denoted $g2$=N, uses no information from industry reports, while likelihood in equation \ref{likelihood1} uses industry information of aggregate market share and is denoted by $g2$=Y. By comparing results across both likelihood functions, we draw attention to whether there is value in using industry report, if available. We measure the model performance using standard metrics: Root Mean Squared Error (RMSE) and symmetric Mean Absolute Percentage Error (sMAPE). In the formulation below sMAPE lies between 0\% and 100\%.\\

\begin{align*}
    \text{RMSE} = \sqrt{\frac{1}{n}\sum_{1}^n{\left(\text{estimate}-\text{actual}\right)}^2};  
    \text{sMAPE} = \frac{100\%}{n}\sum_{1}^n\frac{\left|\text{estimate}-\text{actual}\right|}{\left|\text{estimate}\right|+\left|\text{actual}\right|} 
\end{align*}

\subsection{Interim Evaluation}
\begin{figure}[!h]
\centerline{\includegraphics[width=0.6\columnwidth ]{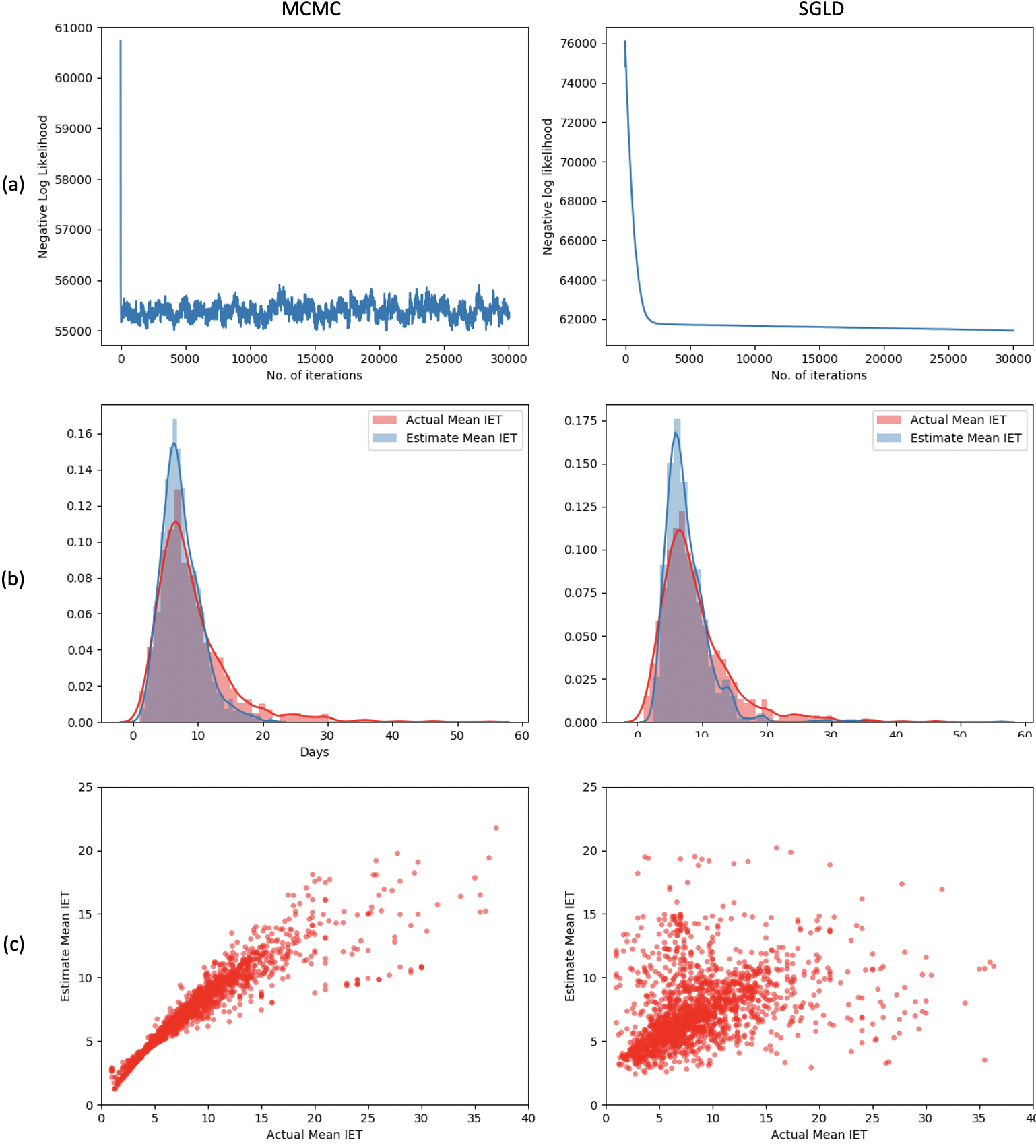}}
\caption{Experiment 1 - Using Erlang-2 distribution, for MCMC and SGLD (a) Negative log likelihood versus iterations; (b) Histogram of IET; (c) Scatter plot for Actual and Estimated Average IET on the site. Histograms show good match between actual and estimated IET distributions from both MCMC and SGLD. Scatter plots show overall good alignment among individual average actual and estimated IET from MCMC; and more scattered for SGLD.}
\label{main_full}
\end{figure}

The purpose of this interim evaluation is to check the assumption of Erlang distribution as a fit for IET. From the estimated parameters ($\beta_i$,\hspace{1mm}$\phi_i$,\hspace{1mm}$\lambda_i$) of the model, we can compute the IET distribution $g_{1i}(.)$ for the focal site, for each $i$. The expectation coming from $g_{1i}(.)$ is the estimated average IET for visits to the focal site for $i$-th user. The actual average IET for the focal site for visits by $i$-th user is known since those visits are observed in the site's own data. The comparison of the estimated and actual IETs on focal site, for all $i$, gives an \textit{interim evaluation} of the modeling framework, by testing the IET estimates and the Erlang assumption. Figure~\ref{main_full} shows the variation of negative log likelihood with iteration, histograms and scatter plots for estimated and actual average IET on focal site, using Erlang-2. The plots for Erlang-1 are similar and are not shown. For both MCMC and SGLD, the negative log likelihood plots show good convergence and the histograms show a good match between distributions of actual and estimated IET. The likelihood convergence plot for SGLD is smoother because it does not use the accept-reject criterion of MCMC. Importantly, individual-level average IETs align well on the scatter plots (scaling across X and Y axes are different) for MCMC. The alignment is better for small values of average IET than for large values. Large (small) values of average IET imply infrequent (frequent) visits, and hence less (more) data points per individual, influencing estimates adversely (favorably). The scatter plots for SGLD are less aligned relative to MCMC, indicating the estimated average IETs are relatively more dispersed.
% The alignment is better for \textit{elec} as the data come from six products with similar visit cycles. However, the large set of products in \textit{full} data represents disparate visit cycles. Figures for likelihood equation \ref{likelihood2} are similar and are not shown.

Coming to the model's performance for IET, Table~\ref{originalResults} shows results for both likelihood equations \ref{likelihood1} ($g2$=Y indicating presence of $g2$) and \ref{likelihood2} ($g2$=N indicating absence of $g2$). Considering MCMC, the values of RMSE are low for both $g2$=Y and $g2$=N; and three values of sMAPE are small (3.39\% to 3.77\%), where sMAPE ranges from 0\% to 100\%. It is higher (12.47\%) for Erlang-1 with $g2$=N. Overall, the results strongly indicate the viability of the proposed model. With the use of SGLD, the RMSE and sMAPE values are about twice that of MCMC. Except for MCMC with Erlang-1, when industry level metric on aggregate market share is absent, that is, $g2$=N, the results in RMSE and sMAPE values, are comparable to using such data, that is, $g2$=Y. The results indicate the potential that the modeling framework is not particularly dependent on the availability of such industry reports. In validation experiments, described next, we throw more light on this issue.

\iffalse
\subsection{Validation Strategy}
Even when our model can estimate PSE, validating the proposed approach is difficult, since a focal site cannot typically access log data of other sites.
%without violating ethics and privacy of users. 
As a novel contribution, we introduce a validation strategy by using only the focal site's data to construct a simulated truth. This is a fairly general validation approach and can be used in situations the site faces, where data about users' engagement on other sites are not available. Consider the dataset of the focal site itself as the truth, 
%which is divided, for each user, into visits to focal sites and to that of other sites, 
where for each user, some visits are randomly suppressed. The \textit{non-suppressed} visits are treated as \textit{observed engagements} to the focal site, and the \textit{suppressed} visits are \textit{unobserved engagements}, or, visits to other sites. The model uses only non-suppressed visits to estimate PSE and IET. Thus, we construct simulated truth from real data of the site, to mimic real-life condition where visit-level data on other sites are not available to the focal site. Actual PSE for each user is obtainable as: (\# visits in non-suppressed data / \# visits in non-suppressed and suppressed data), and actual mean IET across all sites as the average of IET across all visits in non-suppressed and suppressed data.
\fi

\subsection{Validation Strategy}
Even when our model can estimate PSE, validating the proposed approach is difficult, since a focal site cannot typically access log data of other sites.
%without violating ethics and privacy of users. 
As a novel contribution, we introduce a validation strategy by using only the focal site's data to construct a simulated truth. This is a fairly general validation approach and can be used in situations the site faces, where data about users' engagement on other sites are not available. Consider the dataset of the focal site itself as the truth, 
%which is divided, for each user, into visits to focal sites and to that of other sites, 
where for each user, some visits are randomly suppressed. The \textit{non-suppressed} visits are treated as \textit{observed engagements} to the focal site, and the \textit{suppressed} visits are \textit{unobserved engagements}, or, visits to other sites. The model uses only non-suppressed visits to estimate PSE and IET. Thus, we construct simulated truth from real data of the site, to mimic real-life condition where visit-level data on other sites are not available to the focal site. Actual PSE for each user is obtainable as: (\# visits in non-suppressed data / \# visits in non-suppressed and suppressed data), and actual mean IET across all sites as the average of IET across all visits in non-suppressed and suppressed data.

\begin{table*}[!htbp]
\begin{center}
\begin{tabular}{ cccc cc c cc c cc c cc }
    \toprule
    \multirow{2}{*}{\textbf{g2}} & \multirow{2}{*}{\textbf{Set}} & \multirow{2}{*}{$n$} & \multirow{2}{*}{} & \multicolumn{2}{c}{\textbf{MCMC Erlang-2}} && \multicolumn{2}{c}{\textbf{MCMC Erlang-1}} && \multicolumn{2}{c}{\textbf{SGLD Erlang-2}} && \multicolumn{2}{c}{\textbf{SGLD Erlang-1}}\\
    \cmidrule{5-6}\cmidrule{8-9}\cmidrule{11-12}\cmidrule{14-15}
    & & & & \textbf{RMSE} & \textbf{sMAPE} && \textbf{RMSE} & \textbf{sMAPE} && \textbf{RMSE} & \textbf{sMAPE} && \textbf{RMSE} & \textbf{sMAPE}\\
    \midrule
    \multirow{5}{*}{Y} & $all$ & 1243 && 2.99 & 14.48\% && 2.80 & 12.88\% && 6.21 & 12.56\% && 9.75 & 14.91\% \\[0.3ex]
    \cdashline{2-15}\noalign{\vskip 0.6ex}
    & $Q1$ & 310 && 4.61 & 20.90\% && 4.38 & 18.78\% && 11.80 & 15.39\% && 18.92 & 15.64\% \\
    & $Q2$ & 311 && 3.25 & 16.65\% && 2.95 & 13.98\% && 3.15 & 12.69\%  && 3.50 & 13.32\% \\
    & $Q3$ & 311 && 1.74 & 11.41\% && 1.58 & 9.80\% && 1.73 & 9.47\% && 2.37 & 12.23\% \\
    & $Q4$ & 311 && 1.01 & 8.99\% && 0.99 & 8.99\% && 1.39 & 12.67\% && 2.11 & 18.46\% \\[0.3ex]
    \cdashline{1-15}\noalign{\vskip 1ex}
    \multirow{5}{*}{N} & $all$ & 1243 && 3.00 & 14.57\% && 3.63 & 16.15\% && 3.96 & 21.86\% && 3.49 & 13.25\% \\[0.3ex]
    \cdashline{2-15}\noalign{\vskip 0.6ex}
    & $Q1$ & 310 && 4.60 & 20.95\% && 5.32 & 21.09\% && 5.74 & 25.43\% && 5.38 & 14.52\% \\
    & $Q2$ & 311 && 3.24 & 16.62\% && 3.85 & 16.85\% && 4.45 & 24.43\% && 2.50 & 11.41\% \\
    & $Q3$ & 311 && 1.77 & 11.54\% && 2.59 & 13.80\% && 2.80 & 21.99\% && 2.33 & 9.87\% \\
    & $Q4$ & 311 && 1.04 & 9.20\% && 1.68 & 12.89\% && 1.47 & 15.59\% && 2.86 & 17.21\% \\
    \bottomrule
\end{tabular}
\end{center}
\caption{Experiment 2 - Validation of IET, using $60\%$ random suppression of visits per user. For MCMC Erlang-2, RMSE and sMAPE values are similar for g2=Y and g2=N. For MCMC Erlang-1, these values are slightly lower when g2=Y than g2=N. For SGLD Erlang-2, g2=Y performs better than g2=N; for SGLD Erlang-1, g2=Y performs slightly worse than g2=N.}
\label{simulation60Results}
\end{table*}

%Coming to the model's performance for IET, Table~\ref{originalResults} shows results for both likelihood equations \ref{likelihood1} ($g2$=Y indicating presence of $g2$) and \ref{likelihood2} ($g2$=N indicating absence of $g2$). Considering MCMC, the values of RMSE are low for both $g2$=Y and $g2$=N; and three values of sMAPE are small (3.39\% to 3.77\%), where sMAPE ranges from 0\% to 100\%. It is higher (12.47\%) for Erlang-1 with $g2$=N. Overall, the results strongly indicate the viability of the proposed model. With the use of SGLD, the RMSE and sMAPE values are about twice that of MCMC. Except for MCMC with Erlang-1, when industry level metric on aggregate market share is absent, that is, $g2$=N, the results in RMSE and sMAPE values, are comparable to using such data, that is, $g2$=Y. The results indicate the potential that the modeling framework is not particularly dependent on the availability of such industry reports. In validation experiments, described next, we throw more light on this issue.

\begin{figure}[H]
\centerline{\includegraphics[width=5cm]{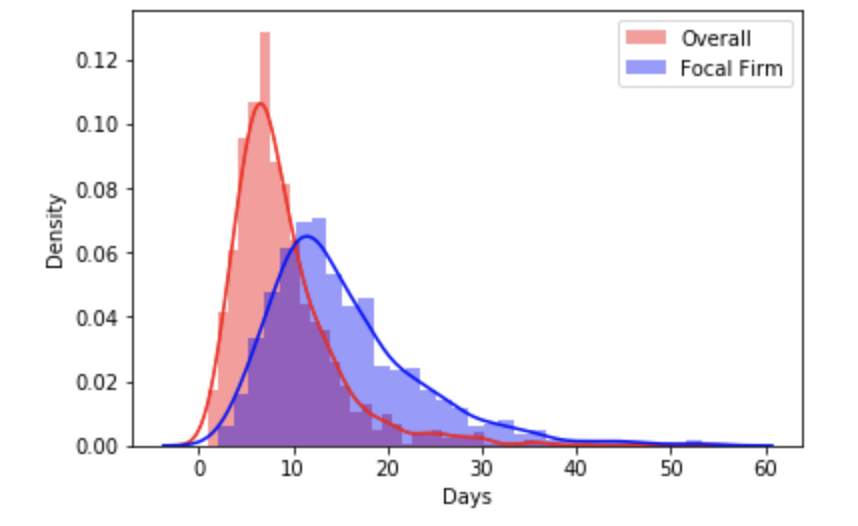}}
\caption{Validation: Empirical distribution of observed IET in all, simulated truth data (All = Focal site + Other sites) vs. that in non-suppressed data (Focal site). The IET distribution for the Focal site is considerably different from that for the All sites.}  
\label{distribution}
\end{figure}

The model estimation mimics the reality by 
%using observed engagements on the focal site (non-suppressed visits), but 
not using unobserved engagements (suppressed visits). Based on non-suppressed visits alone (observed engagements), we estimate IET for all visits and PSE, for each user. To evaluate the model, we compare the estimated PSE and IET with the actual PSE and IET of the simulated data. To ensure a minimum amount of visits per user, we confine to data of users with at least $3$ observed visits after suppression, i.e, $m_i \ge 3, \hspace{1mm} \forall i$. That changes the number of users in the data from $1750$ in experiment 1, to fewer numbers in the following experiments. For robustness, two different experiments are run to validate IET and PSE individually.

\begin{table*}[!htbp]
\begin{center}
\resizebox{\columnwidth}{!}{\begin{tabular}{ cccc cc c cc c cc c cc }
    \toprule
    \multirow{3}{*}{\textbf{g2}} & \multirow{3}{*}{\textbf{Prop-sup}} & \multirow{3}{*}{$n$} & \multirow{3}{*}{} & \multicolumn{2}{c}{\textbf{MCMC Erlang-2}} && \multicolumn{2}{c}{\textbf{MCMC Erlang-1}} && \multicolumn{2}{c}{\textbf{SGLD Erlang-2}} && \multicolumn{2}{c}{\textbf{SGLD Erlang-1}}\\
    \cmidrule{5-6}\cmidrule{8-9}\cmidrule{11-12}\cmidrule{14-15}
    & & & & \textbf{RMSE} & \textbf{sMAPE} && \textbf{RMSE} & \textbf{sMAPE} && \textbf{RMSE} & \textbf{sMAPE} && \textbf{RMSE} & \textbf{sMAPE} \\
    \midrule
    \multirow{5}{*}{Y} & $[0.55, 0.75]$ & 487 && 0.102 & 11.16\% && 0.086 & 9.68\% && 0.206 & 22.12\% && 0.193 & 20.93\% \\[0.3ex]
    \cdashline{2-15}\noalign{\vskip 0.6ex}
    & $[0.55, 0.60]$ & 129 && 0.06 & 5.93\% && 0.022 & 2.07\% && 0.164 & 15.67\% && 0.153 & 14.66\% \\
    & $[0.60, 0.65]$ & 117 && 0.078 & 8.17\% && 0.052 & 5.98\% && 0.193 & 19.74\% && 0.183 & 18.64\% \\
    & $[0.65, 0.70]$ & 146 && 0.107 & 12.39\% && 0.096 & 12.52\% && 0.219 & 24.46\% && 0.205 & 23.2\% \\
    & $[0.70, 0.75]$ & 95 && 0.152 & 20.05\% && 0.141 & 20.17\% && 0.247 & 30.19\% && 0.232 & 28.78\% \\[0.3ex]
    \cdashline{1-15}\noalign{\vskip 1ex}
    \multirow{5}{*}{N} & $[0.55, 0.75]$ & 487 && 0.126 & 13.69\% && 0.157 & 17.48\% && 0.047 & 5.25\% && 0.148 & 16.74\% \\[0.3ex]
    \cdashline{2-15}\noalign{\vskip 0.6ex}
    & $[0.55, 0.60]$ & 129 && 0.071 & 7.19\% && 0.080 & 8.09\% && 0.063 & 7.37\% && 0.109 & 10.57\% \\
    & $[0.60, 0.65]$ & 117 && 0.1 & 10.77\% && 0.126 & 13.95\% && 0.035 & 4.33\% && 0.136 & 14.63\% \\
    & $[0.65, 0.70]$ & 146 && 0.134 & 15.51\% && 0.179 & 21.38\% && 0.029 & 2.87\% && 0.160 & 19.01\% \\
    & $[0.70, 0.75]$ & 95 && 0.182 & 23.33\% && 0.221 & 28.57\% && 0.054 & 7.16\% && 0.185 & 24.22\% \\
    \bottomrule
\end{tabular}}
\end{center}
\caption{Experiments 3 - Validation of PSE, with proportion of suppression of visits per user, \textit{Prop-sup}, randomly drawn from $U[0.55, 0.75]$. The RMSE and sMAPE values are reasonably low, especially for buckets with lower \textit{Prop-sup} in the range $U[0.55, 0.65]$, for all cases of IET distribution and for both g2=Y and g2=N. MCMC yields better performance in terms of RMSE and sMAPE relative to SGLD, except for Erlang-2 with g2=N, where SGLD fares better.}
\label{simulationUniformResults}
\end{table*}

\subsection{Validation of IET}
In experiment 2, we randomly suppress $60\%$ of the visits for each user, resulting in a simulated truth PSE of $0.4$. As shown in Figure~\ref{distribution}, the IET distribution for the simulated truth labeled All = (focal site + other sites) is considerably different from the IET distr for the focal site. This difference found in our validation data is consistent with the marketplace because the engagement behaviors on a focal site is likely to be different from the engagement behaviors across all sites, comprising the focal site and the other sites. Thus, our validation approach reflects the reality of the marketplace. Moreover, this difference is also consistent with our modeling framework, where the IET distribution $f$ for all sites is different from that of the focal site $g1$. 

We estimate using both objective functions, $g2$=Y (when reliable industry-level information for aggregate market share is available), and $g2$=N (when such industry level information is not available). We report evaluations for IET using two different IET distributions, namely Erlang-2 and Erlang1. For further analysis of the effect of number of visits on model performance, we divide users into 4 quartiles $Q1$ to $Q4$, (labeled, \textit{Set}) based on the total number of visits. The number of visits per user increases from $Q1$ to $Q4$.

Results in Table~\ref{simulation60Results} show model performance in estimating IET as compared to simulated truth. The RMSE values are low and sMAPE values are reasonable as well, with most sMAPE values across all comparisons below 15\%. Comparing across $Q1$ to $Q4$, it is evident from the decreasing trends for IET evaluation that the model accuracy increases with the number of visits. Coming to specific comparisons, under MCMC, when using Erlang-2 for IET distribution, performance metric values, RMSE and sMAPE, are similar across $g2$=Y and $g2$=N. With Erlang-1 distribution, RMSE and sMAPE values are slightly lower relative to Erlang-2, when $g2$=Y; however, they are slightly higher when $g2$=N. Also, for MCMC with Erlang-1, RMSE and sMAPE values are higher for $g2$=N relative to that of $g2$=Y. Within SGLD, for $g2$=Y, the performance of Erlang-2 is better than Erlang-1, while for $g2$=N it is poorer. Overall, when $g2$=Y, across $Q1$ to $Q4$, comparison between MCMC and SGLD shows, for Erlang-2 MCMC yields slightly worse performance than SGLD, but MCMC performs equally with SGLD for Erlang-1. When $g2$=N, across $Q1$ to $Q4$, comparison between MCMC and SGLD shows, for Erlang-2 MCMC yields better performance than SGLD, but MCMC performs worse than SGLD for Erlang-1. Hence, these results show that the use of industry information to inform parameter estimation ($g2$=Y) versus not to use ($g2$=N), is dependent on specific IET distribution and algorithm. That said, the performance metrics sMAPE and RMSE indicate acceptable error rate, when $g2$=N as compared to when $g2$=Y, for the comparisons. One intuition for the difference is that estimation of the model parameters benefits more from the additional information ($g2$=Y) when Erlang-1 is used since it has the statistical property of being memoryless, while Erlang-2 is not. Testing this is a valuable research task going forward.

% Erlang-1 recovers the ground truth IET more accurately than Erlang-2, when industry information is used to inform parameter estimation ($g2$=Y). When such industry information is not used ($g2$=N), Erlang-2 returns better performance. One potential explanation for the difference is that estimation of the model parameters benefits more from the additional information ($g2$=Y) when Erlang-1 is used since it has the statistical property of being memoryless, while Erlang-2 is not memoryless. Testing this is a valuable research task going forward.

\subsection{Validation of PSE}
In additional experiment, experiment 3, we validate our estimation of PSE, the focus of this work. For each user, the proportion of visits suppressed (labeled, \textit{Prop-sup}) is sampled randomly from $U[0.55, 0.75]$ and we use the unsuppressed visits for estimation. That is, the proportion of visits representing observed engagements (or actual PSE) lie in $(0.25-0.45)$, across users.

Table~\ref{simulationUniformResults} shows results on different user groups based on \textit{Prop-sup}. More than half of the RMSE values, 22 out of 40 cells, are less than 0.15, indicating good overall performance in recovering PSE. For sMAPE, 21 out of 40 cells ($>$50\%), have values less than 15\%, and 12 out of 40 cells show values less than 10\%. Thus, we find good support in overall performance of our model in recovering PSE. The sMAPE values area bit high \textit{Prop-sup} in $(0.70-0.75)$, a group which uses only $0.25$-$0.30$ proportion of observed engagements for estimation. This proportion of visits can be adequate if it covers many visits, but is not the case in our 4-month data. Improved model performance occurs with lower \textit{Prop-sup} and more data points per user for estimation. More specifically, under MCMC, for buckets with lower \textit{Prop-sup}, in the ranges $0.55$ to $0.60$ and $0.60$ to $0.65$, RMSE and sMAPE values are low. Under MCMC, Erlang-1 performs better when $g2$=Y; but Erlang-2 performs better when $g2$=N. This could be because the dataset suffers from heterogeneity in visit cycles, being a mix of all categories. This issue is worthy of exploring in future research, by examining adequate data for a single category. Under SGLD, Erlang-2 with $g2$=N, returns better performance than the other three combinations. A head to head like comparison between MCMC and SGLD finds that MCMC yields better performance in terms of RMSE and sMAPE relative to SGLD, except for Erlang-2 with $g2$=N, where SGLD fares better. Overall, the results indicate recovery of simulated truth PSE is achieved with reasonable accuracy, especially when there are higher numbers of visits to the focal site (lower \textit{Prop-sup}). 
%While we Erlang-2 or Erlang-1 can be determined with specific data at hand for a specific site. In addition, even when industry-level aggregate market share information is not available ($g2$=N), the model performance achieves good accuracy. 

\section{Conclusions}
%Modeling of users' behaviors is germane to data mining research in ML, spanning search, recommendations, targeting, etc.~\cite{hidasi2015session,elkahky2015multi,zheng2016neural,covington2016deep,hiemstra2021advances, vardasbi2020inverse,karatzoglou2013learning,chen2016learning, lee2010improving}. 
User behavior modeling is di riguer in ML. Yet, our problem is unattended. By using its users' behavior log data on its own site, a focal firm can find great value in figuring out the same users' behaviors on other firm's sites. The behaviors on other sites are however unobservable to the focal firm. The research interest lies in learning individual user level PSE. In learning PSE, we (i) model the research problem of learning to infer unobserved behaviors from the firm's own observed behavior data; and (ii) introduce an evaluation approach within the observed behavior data, since ground truth unobserved behaviors are not known. We estimate model parameters for each individual user through a Hierarchical Bayes approach.
%To address the first aspect, a two-part model using Erlang distribution for inter-engagement time and Markov chain for apportioning engagements between the focal firm and the other firms is proposed. A Hierarchical Bayes approach is used for computing PSEs at the individual level, although for most users only few data points are available. Two methods of Bayesian estimation are shown - MCMC and SGLD. To address the issue of validation stated in the second aspect, we introduce a novelty for log data, in the form of a simulated ground truth based on the focal firm's real data. This simulated ground truth affords a general validation strategy for this common premise of lack of other firms' data, which any online firm faces. 
We show results for two different IET distributions, and two scenarios - one, when the focal firm has no access to reliable industry information of aggregate market share, and two, when the focal firm has this information. This comparison reaffirms that without industry level aggregate market share the model works well; \textit{if} the firm has this data somewhat better results may follow. Note that we do not use any outside data to build extra user level features, but stay within the log data. 
We show our model's performance on this simulated ground truth across two large experiments and find good support for our modeling approach. 
%To achieve the promise of truly personalized marketing, PSE can be an essential metric to figure out how much of each user’s engagement is lost to other firms and thus the potential opportunity that remains to be attained from each user. 

%Even with sparsity of features, shorter duration of data for observed visits and fewer number of users, our model obtains good performance. 
%Our experiments suggest that better performance can ensue with more data on each of these counts. 
We note that the approach extends to news media, social sites, financial services, etc. since the only data used is log data of the focal firm. 
%In another extension, the model can be used to predict the epoch of each user's engagements on other firms' sites. The focal firm can then send messages to the user closer to that epoch. 
Future work can use data from these firms to generalize this learning approach. When firms have user profiling data those can be used as user level features to improve performance. Also, other distributions of IET can apply to different products. In closing, learning to infer unobserved behaviors from observed behavior data is a key area for model development and can benefit from more research attention in ML.

\bibliographystyle{ACM-Reference-Format}
\bibliography{references}

\end{document}